\documentclass[aps,pra,twocolumn,superscriptaddress,showpacs,
nofootonbib]{revtex4}

\usepackage{amssymb}
\usepackage{amsmath}
\usepackage{amsthm}
\usepackage{amsfonts}
\usepackage{color}
\usepackage{amscd}
\usepackage{graphicx}

\newtheorem{thm}{Theorem}%[section]
\newtheorem{defin}{Definition}%[section]
\newtheorem{prop}{Proposition}%[section]
\newcommand{\ket}[1]{|#1\rangle}

\newcommand{\nn}{\nonumber}

\newcommand{\rg}{\mathop{\rm r }\nolimits}
\newcommand{\lin}{\mathop{\rm span }\nolimits}

\begin{document}

\title{Inductive Entanglement Classification of Four Qubits under SLOCC}

\author{L. Lamata}
\email[]{lamata@imaff.cfmac.csic.es} \affiliation{Instituto de
Matem\'{a}ticas y F\'{\i}sica Fundamental, CSIC, Serrano 113-bis,
28006 Madrid, Spain}

\author{J. Le\'{o}n}
\email[]{leon@imaff.cfmac.csic.es} \affiliation{Instituto de
Matem\'{a}ticas y F\'{\i}sica Fundamental, CSIC, Serrano 113-bis,
28006 Madrid, Spain}

\author{D. Salgado}
\email[]{david.salgado@uam.es} \affiliation{Dpto.\ F\'{\i}sica
Te\'{o}rica, Universidad Aut\'{o}noma de Madrid, 28049 Cantoblanco,
Madrid, Spain}

\author{E. Solano}
\email[]{enrique.solano@physik.lmu.de} \affiliation{Physics
Department, ASC, and CeNS, Ludwig-Maximilians-Universit\"at,
Theresienstrasse 37, 80333 Munich, Germany }
\affiliation{Secci\'{o}n F\'{\i}sica, Departamento de Ciencias,
Pontificia Universidad Cat\'{o}lica del Per\'{u}, Apartado Postal
1761, Lima, Peru}
\date{\today}

\begin{abstract}
Using an inductive approach to classify multipartite entangled
states under stochastic local operations and classical communication
introduced recently by the authors [Phys. Rev. A 74, 052336 (2006)],
we give the complete classification of four-qubit entangled pure
states. Apart from the expected degenerate classes, we show that
there exist eight inequivalent ways to entangle four qubits. In this
respect, permutation symmetry is taken into account and states with
a structure differing only by parameters inside a continuous set are
considered to belong to the same class.
\end{abstract}


%\keywords{}

\maketitle

\section{Introduction}

Entanglement is the distinguishing feature of quantum
systems~\cite{Bel87a,Per93a} and is especially useful in using such
systems to process information~\cite{BouEkeZei00a,ChuNie00a}.
Despite its relevance this property is not yet fully understood and
different questions remain open~\cite{HorHorHor01a}. Among them a
complete classification of entangled states under adequately chosen
criteria of equivalence stands as a formidable task. Two of these
criteria appear as outstanding, namely the so-called LU and SLOCC
equivalences. The former stands for Local Unitary, which is
mathematically expressed by the relation $\ket{\psi} \sim_{LU}
\ket{\phi}\Leftrightarrow\ket{\psi}=U^{[1]} \otimes \dots U^{[N]}
\ket{\phi}$ for unitary matrices $U^{[k]}$, whereas the latter
stands for Stochastic Local Operations and Classical Communication,
which is mathematically translated as $\ket{\psi} \sim_{\rm SLOCC}
\ket{\phi} \Leftrightarrow \ket{\psi} = F^{[1]} \otimes \dots
F^{[N]} \ket{\phi}$ for non-singular matrices $F^{[k]}$. Although
these entanglement equivalence definitions can be straightforwardly
generalized to include mixed states, we will restrict ourselves to
the pure state case. Furthermore we will center upon the
identification of entanglement classes under the second criterion.

The SLOCC equivalence (also known as local filtering operations) was
introduced in~\cite{DurVidCir00a} and enjoys a clear physical
motivation: two multipartite pure states $\ket{\psi}$ and
$\ket{\phi}$ are equivalent under SLOCC if one can be obtained from
the other with non-null probability using only local operations upon
each subsystem and classical communication among the different
parties. In this manner, the complete classification of entanglement
for $N=3$ qubits can be derived, namely, any genuinely entangled
three-qubit systems is equivalent either to the $GHZ$ state or to
the $W$ state. The difficulties for more general cases, i.e.\ for
$N\geq 4$, was also pointed out in Ref.~\cite{DurVidCir00a}, where a
non-enumerable set of entanglement classes would be required.

In this work, using our preceding inductive
approach~\cite{LamLeoSalSol06a}, we give the complete classification
of entanglement for $N=4$ qubits. Although this was already
apparently performed in Ref.~\cite{VerDehMooVer02a}, we argue that
their enumeration does not follow the same philosophy as in the
seminal paper by D\"{u}r \emph{et
al.}~\cite{DurVidCir00a,CorTol06a}. In Sec.~\ref{PreRelRes}, we
compile the preceding relevant results which will be later used in
Sec.~\ref{FouQubCla} to obtain in detail all entanglement classes
under SLOCC. Our concluding remarks are presented in
Sec.~\ref{ConRem}.

\section{Review of the Inductive Method}

\label{PreRelRes}

The forthcoming entanglement classification for $N$ qubits is based
on the inductive approach which the present authors introduced in
Ref.~\cite{LamLeoSalSol06a} and in some auxiliary results which were
proved therein. The main idea of this approach was to establish a
classification of the right singular subspace of the coefficient
matrix of each pure state $\ket{\psi}$ expressed in an adequate
product basis (typically the canonical --computational-- basis).
This classification must be carried out according to the
entanglement classes for $N-1$ qubits (hence the term
\emph{inductive}). Consequently, knowing in advance that there are
only two entanglement classes for $2$ qubits, namely the product
class $00$ and the entanglement class $\Psi$, the right singular
subspace $\mathfrak{W}$ of the coefficient matrix of the state of
any three-qubit system results to be of six possible types, i.e.\
two one-dimensional subspaces $\mathfrak{W}=\lin\{\ket{\phi\psi}\}$
and $\mathfrak{W}=\lin\{\ket{\Psi}\}$ driving us to the degenerate
classes $000$ and $0_{1}\Psi_{23}$, respectively, two
two-dimensional product subspaces
$\mathfrak{W}=\ket{\phi}\otimes\mathbb{C}^{2}$ and
$\mathfrak{W}=\mathbb{C}^{2}\otimes\ket{\psi}$, driving us to the
degenerate classes $0_{2}\Psi_{13}$ and $0_{3}\Psi_{12}$,
respectively, and, finally, another two two-dimensional classes
$\mathfrak{W}=\lin\{00,00\}$ and $\mathfrak{W}=\lin\{00,\Psi\}$,
driving us to the well-known $GHZ$ and $W$ genuine entanglement
classes. Let us recall the reader that
$\mathfrak{W}=\lin\{00,\Psi\}$ stands for the fact that the subspace
$\mathfrak{W}$ contains (up to normalization) only one product
vector (i.e.\ only one of type $00$). This will sometimes be also
expressed as $\mathfrak{W}=\lin\{\ket{\phi\varphi},\ket{\Psi}\}$.
This convention will be profusely followed in the next section. We
will employ these six tripartite entanglement classes to find those
of $N=4$ qubits.

Moreover, in the proof, we will make use of the following results,
already proved in~\cite{LamLeoSalSol06a}.

\begin{prop}\label{StruW2}
Let $\mathfrak{W}$ be a two-dimensional subspace in $\mathbb{C}^{2}
\otimes \mathbb{C}^{2}$. Then $\mathfrak{W} =
\lin\{\ket{\phi\varphi},\ket{\Psi}\}$ if, and only if,
$\mathfrak{W}=\lin\{\ket{\phi\varphi},\ket{\phi\bar{\varphi}} +
\ket{\bar{\phi}\varphi}\}$, where\hspace*{1mm}  $\bar{ }$ denotes
linear independence.
\end{prop}

Let us also recall that only two options are left for the structure
of any two-dimensional subspace~\cite{SanTarVid98a}, namely
$\mathfrak{W} = \lin\{00,00\}$ and $\mathfrak{W} = \lin\{00,\Psi\}$,
where the above convention has been used. Complementarily, the
following criterion to discern whether a given tripartite pure state
belongs to any of the six entanglement classes will be also
necessary. We previously need the following definition.

\begin{defin}
Let
\begin{subequations}
\begin{eqnarray}\label{Part1}
C^{(1)}\equiv C_{1|23}&=&\begin{pmatrix}
c_{111}&c_{112}&c_{121}&c_{122}\\
c_{211}&c_{212}&c_{221}&c_{222}
\end{pmatrix},\\\label{Part2}
C^{(2)}\equiv C_{2|13}&=&\begin{pmatrix}
c_{111}&c_{112}&c_{211}&c_{212}\\
c_{121}&c_{122}&c_{221}&c_{222}
\end{pmatrix},\\\label{Part3}
C^{(3)}\equiv C_{3|12}&=&\begin{pmatrix}
c_{111}&c_{121}&c_{211}&c_{221}\\
c_{112}&c_{122}&c_{212}&c_{222}
\end{pmatrix}.
\end{eqnarray}
\end{subequations}

\noindent be the coefficient matrices of the partitions $1|23$,
$2|13$ and $3|12$, respectively, of the tripartite pure state
$\ket{\Psi}=\sum_{ijk}c_{ijk}\ket{i-1}\otimes\ket{j-1} \otimes
\ket{k-1}$. We will define the matrices $W_{1}$ and $W_{2}$ as
\begin{equation}
W_{1}=\begin{pmatrix}
c_{111} & c_{112}\\
c_{121} & c_{122}
\end{pmatrix}\qquad W_{2}=\begin{pmatrix}
c_{211} & c_{212}\\
c_{221} & c_{222}
\end{pmatrix}.
\end{equation}

\end{defin}

With these definitions and remembering that the spectrum of a matrix
$A$ is denoted as $\sigma(A)$, in \cite{LamLeoSalSol06a} we proved
the following

\begin{thm}\label{Clas3Qubit}
Let $\ket{\Psi}$ denote the pure state of a tripartite system and
$C^{(i)}$ its coefficient matrix according to the partition $i|jk$.
Then
\begin{enumerate}
\item $\ket{\Psi}$ belongs to the $000$ class if, and only if,
$\rg(C^{(i)}) = 1$ for all $i=1,2,3$.
\item $\ket{\Psi}$ belongs to the $0_{1} \Psi_{23}$ class if, and
only if, $\rg(C^{(1)})=1$ and $\rg(C^{(k)})=2$ for $k=2,3$.
\item $\ket{\Psi}$ belongs to the $0_{2}\Psi_{13}$ class if, and only
if, $\rg(C^{(2)})=1$ and $\rg(C^{(k)})=2$ for $k=1,3$.
\item $\ket{\Psi}$ belongs to the $0_{3}\Psi_{12}$ class if, and only
if, $\rg(C^{(3)})=1$ and $\rg(C^{(k)})=2$ for $k=1,2$.
\item $\ket{\Psi}$ belongs to the $GHZ$ class if, and only if, one of
the following situations occurs:
\begin{itemize}
\item[i.]  $\rg(C^{(i)})=2$ for all $i=1,2,3$ and $\rg(W_{1}) =
\rg(W_{2}) = 1$.
\item[ii.] $\rg(C^{(i)})=2$ for all $i=1,2,3$, $\rg(W_{j})=2$,
$\rg(W_{k}) = 1$ and $\sigma(W_{j}^{-1}W_{k})$ is non-degenerate.
\item[iii.] $\rg (C^{(i)})=2$ for all $i=1,2,3$, $\rg(W_{1}) = 2$,
$\rg(W_{2}) = 2$ and $\sigma(W_{1}^{-1}W_{2})$ is non-degenerate.
\end{itemize}
\item $\ket{\Psi}$ belongs to the $W$ class if, and only if, one of
the following situations occurs:
\begin{itemize}
\item[i.] $\rg(C^{(i)})=2$ for all $i=1,2,3$, $\rg(W_{j})=2$,
$\rg(W_{k})=1$
and $\sigma(W_{j}^{-1}W_{k})$ is degenerate.
\item[ii.] $\rg(C^{(i)})=2$ for all $i=1,2,3$, $\rg(W_{1})=2$,
$\rg(W_{2})=2$
and $\sigma(W_{1}^{-1}W_{2})$ is degenerate.
\end{itemize}
\end{enumerate}
\end{thm}

\section{Four-qubit classification}

\label{FouQubCla}

The classification inductive procedure for $N=4$ qubits is long,
although systematic. We will firstly detect the degenerate classes,
which furthermore are predictable in advance: $0000$,
$0_{i_{1}}0_{i_{2}}\Psi_{i_{3}i_{4}}$, $0_{i_{1}}GHZ$, $0_{i_{1}}W$
and $\Psi_{i_{1}i_{2}}\Psi_{i_{3}i_{4}}$, with $i_{k}=1,2,3,4$.
Then, we will work out the genuine classes. The classification is
performed according to the structure of the right singular subspace
of the coefficient matrix of the four-partite pure states:
$\mathfrak{W}=\lin\{\Psi_{i},\Psi_{j}\}$, where
$\Psi_{i,j}=000,0_{k}\Psi,GHZ,W$.

\subsection{Degenerate classes}

As stated we will revise in turns each of the possible structures of
the right singular subspace $\mathfrak{W}$ of the coefficient
matrix. Remember that now $\mathfrak{W}\subset
\mathbb{C}^{2}\otimes\mathbb{C}^{2}\otimes\mathbb{C}^{2}$.

\begin{description}
\item[1. $\mathfrak{W}=\lin\{000\}$.-] In this case, we can write
$\mathfrak{W}=\lin\{\ket{\phi\varphi\psi}\}$. Mimicking the same
arguments as in \cite{LamLeoSalSol06a} (see also below for the
genuine classes), we obtain the canonical vector $\ket{0000}$,
corresponding to the degenerate class $0000$.
\item[2. $\mathfrak{W}=\lin\{0_{1}\Psi_{23}\}$.-] This drives us to
the canonical vector $\ket{0000}+\ket{0011}$, corresponding to the
degenerate class $0_{1}0_{2}\Psi_{34}$.
\item[3. $\mathfrak{W}=\lin\{0_{2}\Psi_{13}\}$.-] This drives us to
the canonical vector $\ket{0000}+\ket{0101}$, corresponding to the
degenerate class $0_{1}0_{3}\Psi_{24}$.
\item[4. $\mathfrak{W}=\lin\{0_{3}\Psi_{12}\}$.-] This drives us to
the canonical vector $\ket{0000}+\ket{0110}$, corresponding to the
degenerate class $0_{1}0_{4}\Psi_{23}$.
\item[5. $\mathfrak{W}=\lin\{GHZ\}$.-] This drives us to the canonical
vector $\ket{0000}+\ket{0111}$, corresponding to the degenerate
class $0_{1}GHZ$.
\item[6. $\mathfrak{W}=\lin\{W\}$.-] This drives us to the canonical
vector $\ket{0001}+\ket{0010}+\ket{0100}$, corresponding to the
degenerate class $0_{1}W$.
\item[7. $\mathfrak{W}=\lin\{000,000\}$.-] When $\mathfrak{W} =
\ket{\phi\varphi}\otimes\mathbb{C}^{2}$, we obtain the canonical
vector $\ket{0000}+\ket{1001}$, corresponding to the degenerate
class $0_{2}0_{3}\Psi_{14}$.
\item[8. $\mathfrak{W}=\lin\{000,000\}$.-] When $\mathfrak{W} =
\ket{\phi} \otimes \mathbb{C}^{2}\otimes\ket{\psi}$, we obtain the
canonical vector $\ket{0000}+\ket{1010}$, corresponding to the
degenerate class $0_{2}0_{4}\Psi_{13}$.
\item[9. $\mathfrak{W}=\lin\{000,000\}$.-] When $\mathfrak{W} =
\mathbb{C}^{2} \otimes \ket{\varphi\psi}$, we obtain the canonical
vector $\ket{0000}+\ket{1100}$, corresponding to the degenerate
class $0_{3}0_{4}\Psi_{12}$.
\item[10. $\mathfrak{W}=\lin\{000,000\}$.-] When $\mathfrak{W} =
\ket{\phi}\otimes\lin\{\ket{\varphi\psi},\ket{\bar{\varphi}
\bar{\psi}}\}$, we obtain the canonical vector $\ket{0000} +
\ket{1011}$, corresponding to the degenerate class $0_{2}GHZ$.
\item[11. $\mathfrak{W}=\lin\{000,000\}$.-] When $\mathfrak{W} =
\lin\{\ket{\phi\varphi\psi},\ket{\bar{\phi}\varphi\bar{\psi}}\}$, we
obtain the canonical vector $\ket{0000}+\ket{1101}$, corresponding
to the degenerate class $0_{3}GHZ$.
\item[12. $\mathfrak{W}=\lin\{000,000\}$.-] When $\mathfrak{W} =
\lin\{\ket{\phi\varphi\psi},\ket{\bar{\phi}\bar{\varphi}\psi}\}$, we
obtain the canonical vector $\ket{0000}+\ket{1110}$, corresponding
to the degenerate class $0_{4}GHZ$.
\item[13. $\mathfrak{W}=\lin\{000,0_{1}\Psi_{23}\}$.-] When
$\mathfrak{W} = \ket{\phi} \otimes
\lin\{\ket{\varphi_{1}\psi_{1}},\ket{\bar{\varphi}_{2}
\bar{\psi}_{2}}\}$, we obtain the canonical vector
$\ket{0001}+\ket{0010}+\ket{1000}$, corresponding to the degenerate
class $0_{2}W$.
\item[14. $\mathfrak{W}=\lin\{000,0_{2}\Psi_{13}\}$.-] When
$\mathfrak{W} = \lin\{\ket{\phi_{1} \varphi\psi_{1}} ,
\ket{\phi_{2}\varphi\psi_{2}} +
\ket{\bar{\phi}_{2}\varphi\bar{\psi}_{2}}\}$, we obtain the
canonical vector $\ket{0001}+\ket{0100}+\ket{1000}$, corresponding
to the degenerate class $0_{3}W$.
\item[15. $\mathfrak{W}=\lin\{000,0_{3}\Psi_{12}\}$.-] When
$\mathfrak{W} = \lin\{\ket{\phi_{1}\varphi_{1}} ,
\ket{\phi_{2}\varphi_{2}} + \ket{\bar{\phi}_{2} \bar{\varphi}_{2}}\}
\otimes \ket{\psi}$, we obtain the canonical vector
$\ket{0010}+\ket{0100}+\ket{1000}$, corresponding to the degenerate
class $0_{4}W$.
\item[16. $\mathfrak{W}=\lin\{0_{1}\Psi_{13},0_{1}\Psi_{13}\}$.-]
When $\mathfrak{W}=\mathbb{C}^{2}\otimes\lin\{\ket{\Psi}\}$, we
obtain the canonical vector $\ket{0000} + \ket{0011} + \ket{1100} +
\ket{1111} = \left(\ket{00} + \ket{11}\right) \left(\ket{00} +
\ket{11}\right)$, corresponding to the degenerate class
$\Psi_{12}\Psi_{34}$.
\item[17. $\mathfrak{W} = \lin\{0_{2}\Psi_{13},0_{2}\Psi_{13}\}$.-]
When $\mathfrak{W}=\lin\{\ket{\phi\varphi\psi} +
\ket{\bar{\phi}\varphi\bar{\psi}} , \ket{\phi\bar{\varphi}\psi} +
\ket{\bar{\phi}\bar{\varphi}\bar{\psi}}\}$, we obtain the canonical
vector $\ket{0000} + \ket{0101} + \ket{1010} + \ket{1111} =
\left(\ket{00} + \ket{11}\right)_{13} \left(\ket{00} +
\ket{11}\right)_{24}$, corresponding to the degenerate class
$\Psi_{13}\Psi_{24}$.
\item[18. $\mathfrak{W} = \lin\{0_{3}\Psi_{13},0_{3}\Psi_{13}\}$.-]
When $\mathfrak{W} = \ket{\Psi}\otimes\mathbb{C}^{2}$, we obtain the
canonical vector $\ket{0000} + \ket{0110} + \ket{1001} + \ket{1111}
= \left(\ket{00} + \ket{11}\right)_{14} \left(\ket{00} +
\ket{11}\right)_{23}$, corresponding to the degenerate class
$\Psi_{14}\Psi_{23}$.
\end{description}

No further degenerate classes can be found within the rest of
possible structures for $\mathfrak{W}$. Notice how all of them can
be obtained from the first six and the sixteenth cases by just
permuting the qubit indices. We have preferred to make use of the
systematic approach in order to illustrate its usage.

\subsection{Genuine classes}

In order to be exhaustive we must check out all possible structures
that the right singular subspace $\mathfrak{W}$ may have. An a
priori list is given by $\lin\{000,000\}$,
$\lin\{000,0_{i_{1}}\Psi\}$, $\lin\{000,GHZ\}$, $\lin\{000,W\}$,
$\lin\{0_{i_1}\Psi,0_{i_2}\Psi\}$, $\lin\{0_{i_{1}}\Psi,GHZ\}$,
$\lin\{0_{i_{1}}\Psi,W\}$, $\lin\{GHZ,GHZ\}$, $\lin\{GHZ,W\}$ and
$\lin\{W,W\}$, where $i_{k}=1,2,3$. The same convention as above has
been followed when we say that, e.g., $\lin\{0_{i_{1}}\Psi,GHZ\}$
indicates that only one vector belonging to the class
$0_{i_{1}}\Psi$ is contained in $\mathfrak{W}$, the rest being all
$GHZ$ (and possibly $W$). In other words, in
$\lin\{\Psi_{i},\Psi_{k}\}$ it is understood that $\Psi_{i}$ and
$\Psi_{k}$ are those vectors in $\mathfrak{W}$ with the lowest
degree of entanglement degeneracy, according to the scale $000<
0_{i_{k}}\Psi<GHZ,W$.

\subsubsection{$\mathfrak{W}=\lin\{000,000\}$}

All cases leading to degenerate cases have been considered above.
However, when $\mathfrak{W} = \lin\{\ket{\phi\varphi\psi ,
\bar{\phi} \bar{\varphi} \bar{\psi}}\}$, a well-known entanglement
class arises, namely the $GHZ$ class. We choose non-singular
transformations $F^{[2]}$ so that $\ket{0}=F^{[2]}\ket{\phi}$,
$\ket{1}=F^{[2]}\ket{\bar{\phi}}$ and similarly for $F^{[3,4]}$. We
choose also $F^{[1]}$ so that $[F^{[1]}\ket{\mathbf{v}_{1}}\quad
F^{[1]}\ket{\mathbf{v}_{2}}] =
[\mu_{ij}^{*}]^{-1}\left(\begin{smallmatrix}
\sigma_{1}^{-1}&0\\
0&\sigma_{2}^{-1}\end{smallmatrix}\right)$, where
$\ket{\mathbf{v}_{j}}$ denotes the left singular vectors of the
coefficient matrix $C_{\Phi}$ of the state
$\ket{\Phi}\in\mathbb{C}^{2} \otimes
\mathbb{C}^{2}\otimes\mathbb{C}^{2}\otimes\mathbb{C}^{2}$ and
$\mu_{ij}$ are the coefficients expressing the right singular
vectors $\ket{\mathbf{w}_{k}}$ of $C_{\Phi}$ in terms of
$\ket{\phi\varphi\psi}$ and $\ket{\bar{\phi}
\bar{\varphi}\bar{\psi}}$: $\ket{\mathbf{w}_i} =
\mu_{i1}\ket{\phi\varphi\psi} +
\mu_{i2}\ket{\bar{\phi}\bar{\varphi}\bar{\psi}}$. Then, we can write
\cite{LamLeoSalSol06a}

\begin{widetext}
\begin{eqnarray}
C_{F^{[1]}\otimes\dots \otimes F^{[4]} \ket{\Phi}} & = &
\left(\mu_{ij}^{*}\right)^{-1}\begin{pmatrix}
\sigma_{1}&0\\0&\sigma_{2}
\end{pmatrix}^{-1}\begin{pmatrix}
\sigma_{1}&0\\0&\sigma_{2}
\end{pmatrix}\begin{pmatrix}
\mu_{11}^{*} & 0 & 0 & 0 & 0 & 0 & 0 & \mu_{12}^{*}\\
\mu_{21}^{*} & 0 & 0 & 0 & 0 & 0 & 0 & \mu_{22}^{*}
\end{pmatrix}\nn\\
&=&\begin{pmatrix}
1 & 0 & 0 & 0 & 0 & 0 & 0 & 0\\
0 & 0 & 0 & 0 & 0 & 0 & 0 & 1
\end{pmatrix} .
\end{eqnarray}
\end{widetext}

This matrix corresponds to the canonical state
$\ket{0000}+\ket{1111}$, i.e., to the so-called $GHZ$ state.

\subsubsection{$\mathfrak{W}=\lin\{000,0_{1}\Psi_{23}\}$}

In general, in this case $\mathfrak{W} = \lin\{\ket{\phi_{1}
\varphi_{1} \psi_{1}},\ket{\phi_{2}} \otimes
\left(\ket{\varphi_{2}\psi_{2}} + \ket{\bar{\varphi_{2}}
\bar{\psi}_{2}}\right)\}$. The degenerate case takes place when
$\ket{\phi_{1}}$ and $\ket{\phi_{2}}$ are linearly dependent,
driving us to the class $0_{2}W$, already stated above.

Thus, let us consider the case when $\{\ket{\phi_{1}} ,
\ket{\phi_{2}}\}$ are linearly independent. Now the two-dimensional
subspace $\lin\{\ket{\varphi_{1}\psi_{1}} ,
\ket{\varphi_{2}\psi_{2}} + \ket{\bar{\varphi}_{2}\bar{\psi}_{2}}\}$
can be either of type $\lin\{00,00\}$ or of type $\lin\{00,\Psi\}$:

\begin{enumerate}
\item[i.] When $\lin\{\ket{\varphi_{1}\psi_{1}} , \ket{\varphi_{2}
\psi_{2}} + \ket{\bar{\varphi}_{2}\bar{\psi}_{2}}\}=\lin\{00,00\}$,
then there exist linearly independent vectors $\ket{\phi}$,
$\ket{\bar{\phi}}$, $\ket{\varphi\psi}$ and
$\ket{\bar{\varphi}\bar{\psi}}$ such that
$\ket{\mathbf{w}_{k}}=\mu_{k1}\ket{\phi\varphi\psi} +
\mu_{k2}\ket{\bar{\phi}}\otimes\left(\ket{\varphi\psi} +
\ket{\bar{\varphi}\bar{\psi}}\right)$. Thus, it is immediate to
realize that there exist non-singular transformations $F^{[k]}$,
$k=1,2,3,4$, such that

\begin{widetext}
\begin{eqnarray}
C_{F^{[1]}\otimes\dots\otimes F^{[4]}\ket{\Phi}}&=&\left(\mu_{ij}^{*}
\right)^{-1}\begin{pmatrix}
\sigma_{1}&0\\0&\sigma_{2}
\end{pmatrix}^{-1}\begin{pmatrix}
\sigma_{1}&0\\0&\sigma_{2}
\end{pmatrix}\begin{pmatrix}
\mu_{11}^{*} & 0 & 0 & 0 & \mu_{12}^{*} & 0 & 0 & \mu_{12}^{*}\\
\mu_{21}^{*} & 0 & 0 & 0 & \mu_{22}^{*} & 0 & 0 & \mu_{22}^{*}
\end{pmatrix}\nn\\
&=&\begin{pmatrix}
1 & 0 & 0 & 0 & 0 & 0 & 0 & 0\\
0 & 0 & 0 & 0 & 1 & 0 & 0 & 1
\end{pmatrix}
\end{eqnarray}
\end{widetext}

This canonical coefficient matrix corresponds to the state
$\ket{0000} + \ket{1100} + \ket{1111}$.

\item[ii.] When $\lin\{\ket{\varphi_{1}\psi_{1}} , \ket{\varphi_{2}
\psi_{2}} + \ket{\bar{\varphi}_{2}
\bar{\psi}_{2}}\}=\lin\{00,\Psi\}$, then there exist vectors
$\ket{\phi}$, $\ket{\bar{\phi}}$, $\ket{\varphi}$,
$\ket{\bar{\varphi}}$, $\ket{\psi}$ and $\ket{\bar{\psi}}$ such that
$\ket{\mathbf{w}_{k}}=\mu_{k1} \ket{\phi\varphi\psi} + \mu_{k2}
\ket{\bar{\phi}} \otimes \left(\ket{\varphi\bar{\psi}} +
\ket{\bar{\varphi}\psi}\right)$. Thus, it is immediate to realize
that there exist non-singular transformations $F^{[k]}$,
$k=1,2,3,4$, such that

\begin{widetext}
\begin{eqnarray}
C_{F^{[1]} \otimes \dots \otimes F^{[4]}}\ket{\Phi} & = &
\left(\mu_{ij}^{*}\right)^{-1}\begin{pmatrix}
\sigma_{1}&0\\0&\sigma_{2}
\end{pmatrix}^{-1}\begin{pmatrix}
\sigma_{1}&0\\0&\sigma_{2}
\end{pmatrix}\begin{pmatrix}
\mu_{11}^{*} & 0 & 0 & 0 & 0 & \mu_{12}^{*} & \mu_{12}^{*} & 0\\
\mu_{21}^{*} & 0 & 0 & 0 & 0 & \mu_{22}^{*} & \mu_{22}^{*} & 0
\end{pmatrix}\nn\\
&=&\begin{pmatrix}
1 & 0 & 0 & 0 & 0 & 0 & 0 & 0\\
0 & 0 & 0 & 0 & 0 & 1 & 1 & 0
\end{pmatrix}
\end{eqnarray}
\end{widetext}

This canonical coefficient matrix corresponds to the state
$\ket{0000} + \ket{1101} + \ket{1110}$.
\end{enumerate}

\subsubsection{$\mathfrak{W}=\lin\{000,0_{2}\Psi_{13}\}$}

Although it is always possible to reproduce a similar argument to
the one above, we will resort to permutation symmetry, thus leading
to the degenerate class $0_{3}W$ (already stated) and the genuine
classes with canonical states given by
$\ket{0000}+\ket{1010}+\ket{1111}$ and
$\ket{0000}+\ket{1011}+\ket{1110}$.

\subsubsection{$\mathfrak{W}=\lin\{000,0_{3}\Psi_{12}\}$}

With analogous arguments, in this case the degenerate class is
$0_{4}W$ (already stated) and the genuine classes are identified by
the canonical states given by $\ket{0000}+\ket{1001}+\ket{1111}$ and
$\ket{0000}+\ket{1011}+\ket{1101}$.

\subsubsection{$\mathfrak{W}=\lin\{000,GHZ\}$}

This class is highly richer than those above. In this case
$\mathfrak{W} = \lin\{\ket{\phi_{1}\varphi_{1} \psi_{1}} ,
\ket{\phi_{2} \varphi_{2} \psi_{2}} + \ket{\bar{\phi}_2
\bar{\varphi}_{2} \bar{\psi}_{2}} \}$, with the restriction that no
product state other than $\ket{\phi_{1}\varphi_{1}\psi_{1}}$ and no
$0_{k}\Psi$ state belong to $\mathfrak{W}$ (otherwise we would be in
one of the preceding classes). We define now non-singular
transformations $F^{[k]}$ such that $F^{[2]}\ket{\phi_{2}} =
\ket{0}$, $F^{[2]}\ket{\bar{\phi}_{2}} = \ket{1}$, and similarly for
$\ket{\varphi_{2}}$, $\ket{\bar{\varphi}_{2}}$, $\ket{\psi_{2}}$ and
$\ket{\bar{\psi}_{2}}$. We define also $F^{[1]}$ such that $[F^{[1]}
\ket{\mathbf{v}_{1}} \quad F^{[1]}
\ket{\mathbf{v}_{2}}]=[\mu_{ij}^{*}]^{-1}
\left(\begin{smallmatrix} \sigma_{1}^{-1}&0\\
0&\sigma_{2}^{-1}\end{smallmatrix}\right)$, with the same
conventions as above. Then, the canonical vector can be written
as~\footnote{Irrelevant complex conjugation upon each factor has
been omitted for ease of notation.}

\begin{equation}\label{Gen000GHZ}
\ket{0\phi\varphi\psi}+\ket{1000}+\ket{1111} .
\end{equation}

The factor vectors $\ket{\phi}$, $\ket{\varphi}$ and $\ket{\psi}$
are arbitrary up to the restriction of not producing more than one
$000$ vector and no $0_{k}\Psi$ vector in $\mathfrak{W}$. Although
this recipe embraces all possible cases, we will single out two
different subsets within this class, namely (i) that of states
leading to right singular subspaces $\mathfrak{W}$ with no $W$ state
in it, i.e., containing only one product vector and $GHZ$ vectors
and (ii) that of states leading to right singular subspaces
$\mathfrak{W}$ with at least one $W$ vector in it. Indeed, we will
show that there will be either one or two $W$ vectors in
$\mathfrak{W}$.

\begin{enumerate}
\item[i.] The set of states leading to $\mathfrak{W}$ containing no
$W$ vectors is that identified by the canonical ones given by
$\ket{001\psi} + \ket{1000}+\ket{1111}$,
$\ket{00\varphi1}+\ket{1000}+\ket{1111}$ and
$\ket{0\phi01}+\ket{1000}+\ket{1111}$, with no restrictions upon the
factor vectors. Moreover, if we denote the components of
$\ket{\phi}$ in the (non-necessarily  orthonormal) basis
$\{\ket{\phi_{2}},\ket{\bar{\phi}_{2}}\}$ by $\phi_{0}$ and
$\phi_{1}$ and similarly for $\ket{\varphi}$ and $\ket{\psi}$, then
whenever $0\neq\sqrt{\phi_{0} \varphi_{0}\psi_{0}} =
\pm\sqrt{\phi_{1}\varphi_{1}\psi_{1}}\neq 0$, there will also be a
unique product vector, the rest being all $GHZ$ in the right
singular vector of the generic state~\eqref{Gen000GHZ}.

This is proved resorting to theorem~\ref{Clas3Qubit}. We will
illustrate the calculation with the first canonical vector
$\ket{001\psi}+\ket{1000}+\ket{1111}$, whose right singular subspace
is spanned by $\ket{01\psi}$ and
$\frac{1}{\sqrt{2}}\ket{000}+\frac{1}{\sqrt{2}}\ket{111}$. Thus, a
generic vector in $\mathfrak{W}$ will be of the form
$\alpha\left(\ket{000}+\ket{111}\right)+\beta\ket{01\psi}$, whose
coefficient matrix is (cf.\ \eqref{Part1}-\eqref{Part3})

\begin{equation}
C^{(1)}=\begin{pmatrix}
\alpha& 0 & \beta\psi_{0}^{*} & \beta\psi_{1}^{*}\\
0 & 0 & 0 &\alpha
\end{pmatrix}
\end{equation}

For any values of $\psi_{k}^{*}$ it is clear that $\rg(C^{(i)}) = 2$
for $i=1,2,3$ (except for the trivial case $\alpha=0$). Then if
$\psi_{1}=0$, $\rg(W_{1}) = \rg(W_{2}) = 1$, thus it is a $GHZ$
vector; if $\psi_{1}\neq 0$, $\rg(W_{1})=2$ and $\rg(W_{2})=1$ and
the spectrum of $W^{-1}_{1}W_{2}$ is never degenerate. It is again a
$GHZ$ vector for
any value of $\ket{\psi}$.\\

The other canonical vectors are handled in a similar fashion and
they can additionally be obtained under permutations among qubits
$2$, $3$ and $4$. Among them we single out the particular cases
$\ket{0000}+\ket{1101}+\ket{0111}$,
$\ket{0000}+\ket{1011}+\ket{0111}$ and
$\ket{0000}+\ket{1110}+\ket{0111}$, which can also be obtained from
those in the $0_{k}\Psi$ classes by permutations among the
\emph{four} qubits.

For the generic state \eqref{Gen000GHZ}, let us write, as explicitly
supposed,
$\mathfrak{W}=\lin\{\ket{\phi\varphi\psi},\ket{000}+\ket{111}\}$,
with no vector $\ket{\phi}$, $\ket{\varphi}$ and $\ket{\psi}$ equal
to $\ket{0}$ or $\ket{1}$. If we denote again by subindices $0$ and
$1$ the components of each vector $\ket{\phi}$, $\ket{\varphi}$ and
$\ket{\psi}$ in the $\{\ket{0},\ket{1}\}$ basis, then the
coefficient matrix of a generic vector in $\mathfrak{W}$ will be

\begin{equation}\label{MatGen000GHZ}
C=\begin{pmatrix} \alpha\phi_{0}\varphi_{0}\psi_{0}+\beta &
\alpha\phi_{0}\varphi_{0} \psi_{1} &
\alpha\phi_{0}\varphi_{1}\psi_{0} &  \alpha\phi_{0}\varphi_{1}
\psi_{1} \\ \alpha\phi_{1}\varphi_{0}\psi_{0} &
\alpha\phi_{1}\varphi_{0}\psi_{1} &
\alpha\phi_{1}\varphi_{1}\psi_{0} &
\alpha\phi_{1}\varphi_{1}\psi_{1}+\beta
\end{pmatrix}
\end{equation}

\noindent where no coefficient is null, as supposed. Then it should
be clear that $\rg(C^{(i)})=2$ for all $i=1,2,3$. Furthermore, since
$\textrm{det}W_{1}=\alpha\beta\phi_{0}\varphi_{1}\psi_{1}$ and
$\textrm{det}W_{2}=\alpha\beta\phi_{1}\varphi_{0}\psi_{0}$, we will
have $\rg(W_{1})=\rg(W_{2})=2$ (except for the trivial cases
$\alpha\cdot\beta=0$), thus we must check the degeneracy of the
spectrum of $W_{1}^{-1}W_{2}$: $\sigma(W_{1}^{-1}W_{2})$ is
degenerate whenever the discriminant of the characteristic
polynomial is null, i.e.\ whenever
$\left(\textrm{tr}(W_{1}^{-1}W_{2})\right)^{2}-4\textrm{det}
(W_{1}^{-1}W_{2})=0$, which drives us to the condition $\beta +
\alpha\left(\sqrt{\phi_{0}\varphi_{0}\psi_{0}} \pm \sqrt{\phi_{1}
\varphi_{1}\psi_{1}}\right)^{2}=0$, which inmediately yields the
above stated conditions.

\item[ii.] The set of states leading to $\mathfrak{W}$ containing at
least one $W$ vector comprises the rest of possibilites. To prove
this assertion we need to show that all possibilities for
$\mathfrak{W}$ having no $W$ vector has been already considered. The
most generic case occurs when $\mathfrak{W} =
\lin\{\ket{\phi_{1}\varphi_{1}\psi_{1}},\ket{\phi_{2}\varphi_{2}
\psi_{2}} + \ket{\bar{\phi}_{2}\bar{\varphi}_{2}\bar{\psi}_{2}}\}$,
which we detach in the following particular cases:

\begin{itemize}
\item If $\ket{\phi_{1}\varphi_{1}\psi_{1}} = \ket{\phi_{2}
\varphi_{2}
\psi_{2}},\ket{\bar{\phi}_{2}\bar{\varphi}_{2}\bar{\psi}_{2}}$, then
two $000$ vectors belong to $\mathfrak{W}$. This case is ruled out.
\item If $\ket{\phi_{1}\varphi_{1} \psi_{1}} = \ket{\phi_{2}
\varphi_{2} \psi} , \ket{\bar{\phi}_{2}\bar{\varphi}_{2}\psi}$, with
$\ket{\psi}\neq\ket{\psi_{2}},\ket{\bar{\psi}_{2}}$, respectively,
then there will be a $0_{k}\Psi$ in $\mathfrak{W}$ unless
$\ket{\psi}=\ket{\bar{\psi}_{2}},\ket{\psi_{2}}$, respectively. We
give the proof for the first case. Let
$\ket{\phi_{1}\varphi_{1}\psi_{1}}=\ket{\phi_{2}\varphi_{2}\psi}$,
with $\ket{\psi}\neq\ket{\psi_{2}}$. A generic vector in
$\mathfrak{W}$ will be of the form
$\alpha\ket{\phi_{2}\varphi_{2}\psi} + \beta\left(\ket{\phi_{2}
\varphi_{2}\psi_{2}} + \ket{\bar{\phi}_{2}\bar{\varphi}_{2}
\bar{\psi}_{2}}\right) = \ket{\phi_{2} \varphi_{2}}
\left(\alpha\ket{\psi} + \beta\ket{\psi_{2}}\right) +
\beta\ket{\bar{\phi}_{2}\bar{\varphi}_{2}\bar{\psi}_{2}}$. It is
clear that, provided that $\ket{\psi}$ and $\ket{\bar{\psi}_{2}}$
are linearly independent, it is always possible to find
$\alpha,\beta$ such that
$\alpha\ket{\psi}+\beta\ket{\bar{\psi}_{2}}=\bar{\beta}\ket{
\bar{\psi}_{2}}$, driving us to a $0_{3}\Psi$ vector in
$\mathfrak{W}$, against the general hypothesis. By symmetry, all
other cases can be accounted for in a similar fashion, leaving us in
the cases that whenever two vectors in $\ket{\phi\varphi\psi}$ are
equal to two vectors in $\ket{\phi_{2}\varphi_{2}\psi_{2}}$ or
$\ket{\bar{\phi}_{2}\bar{\varphi}_{2}\bar{\psi}_{2}}$, then the
other must be the third one corresponding to the opposite. For
instance, if $\ket{\phi_{2}\varphi_{2}\psi_{2}}=\ket{000}$,
$\ket{\bar{\phi}_{2}\bar{\varphi}_{2}\bar{\psi}_{2}}=\ket{111}$ and
$\ket{\phi_{1}\varphi_{1}\psi_{1}}=\ket{00\psi}$, it must be
$\ket{\psi}=\ket{1}$.
\item If $\ket{\phi_{1}\varphi_{1}\psi_{1}}=\ket{\phi_{2}\varphi
\psi}$, with $\ket{\varphi}\neq\ket{\varphi_{2}}$,
$\ket{\psi}\neq\ket{\psi_{2}}$ and
$\ket{\varphi\psi}\neq\ket{\bar{\varphi}_{2}\bar{\psi}_{2}}$, use
the same non-singular $F^{[k]}$ as in the generic case
\eqref{Gen000GHZ} to arrive at the coefficient matrix for a general
vector in $\mathfrak{W}$

\begin{equation}
\begin{pmatrix}
\alpha\varphi_{0}\psi_{0}+\beta & \alpha\varphi_{0}\psi_{1} & \alpha
\varphi_{1}\psi_{0} & \alpha\varphi_{1}\psi_{1}\\
0 & 0 & 0 & \beta
\end{pmatrix}
\end{equation}

\noindent with the above conditions translated as $\ket{\psi} \neq
\ket{0}\Leftrightarrow\varphi_{1}\neq 0$,
$\ket{\psi}\neq\ket{0}\Leftrightarrow\psi_{1}\neq 0$ and
$\ket{\varphi\psi}\neq\ket{11}\Leftrightarrow\left(\varphi_{0},
\psi_{0}\neq 0\right)$ simultaneously. It is then clear that (except
for the trivial cases $\alpha\cdot\beta=0$) $\rg(C^{(i)})=2$ for all
$i=1,2,3$, $\rg(W_{1})=2$ and $\rg(W_{2})=1$. The spectrum
$\sigma(W_{2}^{-1}W_{1})$ is degenerate only if
$\alpha\varphi_{0}\psi_{0}+\beta=0$. Thus there will be no $W$
vector in $\mathfrak{W}$ only when $\varphi_{0}\psi_{0}=0$, i.e.\
when $\ket{\varphi}=\ket{1}$ or $\ket{\psi}=\ket{1}$, which going
back before the application of the non-singular $F^{[k]}$, means
that $\ket{\varphi_{1}}=\ket{\bar{\varphi}_{2}}$ and
$\ket{\psi_{1}}=\ket{\bar{\psi}_{2}}$. By using $\sigma_{x}$ upon
qubits $2,3,4$ the symmetric case
$\ket{\phi_{1}\varphi_{1}\psi_{1}}=\ket{\bar{\phi}_{2}\varphi\psi}$,
with $\ket{\varphi}\neq\ket{\bar{\varphi}_{2}}$,
$\ket{\psi}\neq\ket{\bar{\psi}_{2}}$ and
$\ket{\varphi\psi}\neq\ket{\varphi_{2}\psi_{2}}$ is also included in
this analysis.  By permutations among the qubits $2,3,4$ all
preceding cases are accounted for in a similar fashion.
\item Finally the case $\ket{\phi_{1}\varphi_{1}\psi_{1}} =
\ket{\phi \varphi\psi}$, with
$\ket{\phi}\neq\ket{\phi_{2}},\ket{\bar{\phi}_{2}}$;
$\ket{\varphi}\neq\ket{\varphi_{2}},\ket{\bar{\varphi}_{2}}$ and
$\ket{\psi}\neq\ket{\psi_{2}},\ket{\bar{\psi}_{2}}$, drives us,
after application of the non-singular $F^{[k]}$, to the coefficient
matrix \eqref{MatGen000GHZ}, already accounted for.
\end{itemize}

No more options are left, thus we have scrutinized all possible
$\mathfrak{W}$'s with structure $\lin\{000,GHZ\}$.

\end{enumerate}

\subsubsection{$\mathfrak{W}=\lin\{000,W\}$}

By $\lin\{000,W\}$ we indicate that only one $000$ vector and $W$
vectors belong to $\mathfrak{W}$. The generic case will be that
expressed by $\mathfrak{W}=\lin\{\ket{\phi_{1}\varphi_{1}
\psi_{1}},\ket{\phi_{2}\varphi_{2}\bar{\psi}_{2}} +
\ket{\phi_{2}\bar{\varphi}_{2}\psi_{2}}+\ket{\bar{\phi}_{2}
\varphi_{2} \psi_{2}}\}$. We will show that there will be no $GHZ$
state in $\mathfrak{W}$ only if $\ket{\phi_{1} \varphi_{1} \psi_{1}}
= \ket{\phi_{2}\varphi_{2}\psi_{2}}$, which after application of the
non-singular transformations $F^{[k]}$ will drive us to the
canonical state $\ket{0001}+\ket{0010}+\ket{0100}+\ket{1000}$, i.e.\
the $W$ state for $4$ qubits. As before, we will be exhaustive:

\begin{itemize}
\item If $\ket{\phi_{1}\varphi_{1}\psi_{1}} = \ket{\phi_{2}
\varphi_{2} \bar{\psi}_{2}},\ket{\phi_{2}\bar{\varphi}_{2}\psi_{2}},
\ket{\bar{\phi}_{2}\varphi_{2}\psi_{2}}$, then it is clear that a
$0_{k}\Psi$ can be found in $\mathfrak{W}$ against the general
hypothesis. This case is then ruled out.
\item If $\ket{\phi_{1}\varphi_{1}\psi_{1}} = \ket{\phi_{2}
\varphi_{2} \psi}$, with $\ket{\psi}\neq\ket{\bar{\psi}_{2}}$, after
application of the non-singular $F^{[k]}$, we obtain the coefficient
matrix for a generic vector in $\mathfrak{W}$
\begin{equation}\label{000W}
\begin{pmatrix}
\alpha\psi_{0} & \alpha\psi_{1}+\beta & \beta & 0\\
\beta & 0 & 0 & 0
\end{pmatrix}
\end{equation}

If $\alpha\psi_{1}+\beta=0$, then there will exist non-null
$\alpha,\beta$ such that a $0_{3}\Psi$ vector will belong to
$\mathfrak{W}$, against the hypothesis. Thus it must be
$\psi_{1}=0$, i.e.\ $\ket{\psi}=\ket{0}$. Under this restriction and
after using theorem \ref{Clas3Qubit} upon the coefficient matrix
\eqref{000W}, all vectors will be of type $W$, since
$\rg(C^{(i)})=2$ for all $i=1,2,3$, $\rg(W_{1})=2$, $\rg(W_{2})=1$
and the spectrum $\sigma(W^{-1}_{1}W_{2})$ is always degenerate. By
permutation symmetry the cases
$\ket{\phi_{1}\varphi_{1}\psi_{1}}=\ket{\phi_{2}\varphi\psi_{2}},
\ket{\phi\varphi_{2}\psi_{2}}$, with the corresponding restrictions,
are also considered in this analysis.
\item If $\ket{\phi_{1}\varphi_{1}\psi_{1}} = \ket{\phi_{2}\varphi
\bar{\psi}_{2}}$, a similar argument leads to the coefficient matrix
for a generic vector in $\mathfrak{W}$ given by

\begin{equation}\label{000W2}
\begin{pmatrix}
0 & \alpha\varphi_{0}+\beta & \beta & 0\alpha\varphi_{1}\\
\beta & 0 & 0 & 0
\end{pmatrix}
\end{equation}

After application of theorem \ref{Clas3Qubit}, it should be clear
that it is always possible to find non-null $\alpha,\beta$ such that
\eqref{000W2} corresponds to a $GHZ$ vector in $\mathfrak{W}$,
against the hypothesis. By permutation symmetry, the rest of cases
with the $000$ generator having two factors in common with some of
components of the $W$ generator is also contained in this analysis.
\item If $\ket{\phi_{1}\varphi_{1}\psi_{1}}=\ket{\phi_{1}
\varphi\psi}$, with $\ket{\varphi} \neq \ket{\varphi_{2} ,
\bar{\varphi}_{2}}$ and $\ket{\psi} \neq \ket{\psi_{2} ,
\bar{\psi}_{2}}$, after a similar argument the coefficient matrix of
a generic vector in the right singular subspace will be

\begin{equation}
\begin{pmatrix}
\alpha\varphi_{0}\psi_{0} & \alpha\varphi_{0}\psi_{1} + \beta &
\alpha \varphi_{1}\psi_{0}+\beta & \alpha\varphi_{1}\psi_{1}\\
0 & 0 & 0 & \beta
\end{pmatrix}
\end{equation}

It should be clear that $\rg(C^{(i)}) = 2$ for all $i=1,2,3$. Now if
$\varphi_{1}\psi_{0}+\varphi_{0}\psi_{1}\neq 0$, there will always
exist non-null $\alpha,\beta$ such that $\rg(W_{1})=\rg(W_{2})=1$,
hence a $GHZ$ vector, against the hypothesis. On the contrary, if
$\varphi_{1}\psi_{0}+\varphi_{0}\psi_{1}= 0$, then $\rg(W_{1})=2$
and $\rg(W_{2})=1$, but it is always possible to find non-null
$\alpha,\beta$ such that $\sigma(W_{1}^{-1}W_{2})$ is
non-degenerate, hence $GHZ\in\mathfrak{W}$. Notice that at most
another $W$ vector, apart from the generator, can be found. By
permutation symmetry and by using $\sigma_{x}$ upon each qubit, the
rest of cases in which the $000$ generator contains one common
factor vector with the $W$ generator is also considered in this
analysis.
\item Finally, if $\ket{\phi_{1}\varphi_{1}\psi_{1}} =
\ket{\phi\varphi \psi}$, with no common factor vector with the $W$
generator, a similar argument leads to the coefficient matrix for a
generic vector in the right singular subspace given by

\begin{equation}
\begin{pmatrix}
\alpha\phi_{0}\varphi_{0} \psi_{0} & \alpha\phi_{0} \varphi_{0}
\psi_{1} + \beta&
\alpha\phi_{0}\varphi_{1}\psi_{0}+\beta&\alpha\phi_{0}\varphi_{1}
\psi_{1}\\ \alpha\phi_{1}\varphi_{0}\psi_{0}+\beta
&\alpha\phi_{1}\varphi_{0}\psi_{1} &
\alpha\phi_{1}\varphi_{1}\psi_{0}&\alpha\phi_{1}\varphi_{1}\psi_{1}
\end{pmatrix}
\end{equation}

A systematic application of theorem~\ref{Clas3Qubit} makes clear
that there always exist $\alpha,\beta$ such that this matrix
corresponds to a $GHZ$ state, against the hypothesis.

\end{itemize}

\subsubsection{$\mathfrak{W}=\lin\{0_{1}\Psi_{23},0_{1}\Psi_{23}\}$}

The generic case will be $\mathfrak{W}=\lin\{\ket{\phi_{1}\Psi_{1}}
, \ket{\phi_{2}\Psi_{2}}\}$, where $\ket{\Psi_{k}}$ denotes a
bipartite entangled vector. Several cases appear:

\begin{enumerate}
\item[i.] When $\ket{\phi_{1}}$ and $\ket{\phi_{2}}$ are linearly
dependent, then $\mathfrak{W} = \ket{\phi} \otimes
\lin\{\ket{\Psi_{1}},\ket{\Psi_{2}}\}$, driving us to the already
known degenerate classes $0_{1}0_{2}\Psi$, $0_{2}$GHZ and $0_{2}W$,
respectively, when it adopts the structures
$\lin\{\ket{\phi\Psi}\}$, $\ket{\phi}\otimes\lin\{00,00\}$ and
$\ket{\phi}\otimes\lin\{00,\Psi\}$. In the forthcoming cases, we
will then impose the linear independence of $\ket{\phi_{1}}$ and
$\ket{\phi_{2}}$.
\item[ii.] If $\ket{\Psi_{1}}$ and $\ket{\Psi_{2}}$ are linearly
dependent, then $\mathfrak{W}=\mathbb{C}^{2}\otimes\ket{\Psi}$,
corresponding to the degenerate class $\Psi_{12}\Psi_{34}$, as
stated above. Therefore, we will also impose the linear independence
of $\ket{\Psi_{1}}$ and $\ket{\Psi_{2}}$.
\item[iii.] When $\lin\{\ket{\Psi_{1}},\ket{\Psi_{2}}\} =
\lin\{00,00\}$, then it is always possible to find linear
independent factor vectors such that $\ket{\Psi_{j}} = a_{j1}
\ket{\varphi\psi} + a_{j2} \ket{\bar{\varphi} \bar{\psi}}$, with
$a_{j1}\cdot a_{j2}\neq 0$. Defining non-singular $F^{[k]}$ such
that $F^{[2]}[a_{11}\ket{\phi_{1}}] = \ket{0}$,
$F^{[2]}[a_{21}\ket{\phi_{2}}] = \ket{1}$, $F^{[3]}[\ket{\varphi}] =
\ket{0}$, $F^{[3]}[\ket{\bar{\varphi}}]=\ket{1}$,
$F^{[4]}[\ket{\psi}]=\ket{0}$, $F^{[4]}[\ket{\bar{\psi}}]=\ket{1}$
and $F^{[1]}$ as in preceding cases, we obtain the canonical matrix

\begin{equation}
\begin{pmatrix}
1 & 0 & 0 & \lambda_{1} & 0 & 0 & 0 & 0\\
0 & 0 & 0 & 0 & 1 & 0 & 0 & \lambda_{2}
\end{pmatrix}
\end{equation}

\noindent where we have defined $\lambda_{i}\equiv\frac{a_{i2}^{*}}
{a_{i1}^{*}}$. This matrix corresponds to the canonical state
$\ket{0000} + \ket{1100} + \lambda_{1}\ket{0011} +
\lambda_{2}\ket{1111}$. with $\lambda_{1}\neq\lambda_{2}$.

\item[iv.] When $\lin\{\ket{\Psi_{1}},\ket{\Psi_{2}}\} = \lin\{00,
\Psi\}$, then it is always possible to find linear independent
factor vectors such that $\ket{\Psi_{j}} = a_{j1} \ket{\varphi\psi}
+
a_{j2}\left(\ket{\varphi\bar{\psi}}+\ket{\bar{\varphi}\psi}\right)$,
with $a_{j1}\cdot a_{j2}\neq 0$. Defining non-singular $F^{[k]}$
such that $F^{[2]}[a_{11}\ket{\phi_{1}}]=\ket{0}$,
$F^{[2]}[a_{21}\ket{\phi_{2}}]=\ket{1}$,
$F^{[3]}[\ket{\varphi}]=\ket{0}$,
$F^{[3]}[\ket{\bar{\varphi}}]=\ket{1}$,
$F^{[4]}[\ket{\psi}]=\ket{0}$, $F^{[4]}[\ket{\bar{\psi}}]=\ket{1}$
and $F^{[1]}$ as in preceding cases, we obtain the canonical matrix

\begin{equation}
\begin{pmatrix}
1 & \lambda_{1} & \lambda_{1} & 0 & 0 & 0 & 0 & 0\\
0 & 0 & 0 & 0 & 1 & \lambda_{2} & \lambda_{2} & 0
\end{pmatrix}
\end{equation}

\noindent where we have defined $\lambda_{i} \equiv
\frac{a_{i2}^{*}}{a_{i1}^{*}}$. This matrix corresponds to the
canonical state $\ket{0000}+\ket{1100} + \lambda_{1} \ket{0001} +
\lambda_{1} \ket{0010} + \lambda_{2} \ket{1101} + \lambda_{2}
\ket{1110}$, with $\lambda_{1}\neq\lambda_{2}$.

\end{enumerate}

All possible cases have been considered. Furthermore, notice that
several states obtained by permutations involving qubit $1$ in
preceding cases are enclosed in this class. This will also happen in
forthcoming classes: some permutations involving qubit $1$ usually
involve a change of entanglement class, although the general
structure of the canonical state shows permutation symmetry
\footnote{This already happened in the tripartite classification:
classes $0_{i_{k}}\Psi$ drive us to different right singular
subspaces, despite the permutation symmetry between the three
qubits. The ultimate reason is the singularization of qubit $1$
produced by the choice of the representation of the coefficient
matrix $C^{(1)}$.}.

\subsubsection{$\mathfrak{W}=\lin\{0_{1}\Psi_{23},0_{2}\Psi_{13}\}$}

The generic case is given by $\mathfrak{W} =
\lin\{\ket{\phi_{1}\Psi_{1}} , \ket{\phi_{2}\varphi_{2}\psi_{2}} +
\ket{\bar{\phi}_{2}\varphi_{2}\bar{\psi}_{2}}\}$. Two possibilities
arise:

\begin{enumerate}
\item[i.] $\lin\{\ket{\Psi_{1}} , \ket{\varphi_{2} \psi_{2}}\} =
\lin\{00,00\} = \lin\{\ket{\varphi_{2} \psi_{2}} ,
\ket{\bar{\varphi}_{2} \bar{\bar{\psi}}_{2}}\}$, where the double
bar also indicates linear independence with respect to
$\ket{\psi_{2}}$, although possible linear dependence with
$\ket{\bar{\psi}_{2}}$. Then a generic vector in $\mathfrak{W}$ will
be given by $\alpha\ket{\phi_{1}}\left(a\ket{\varphi_{2}\psi_{2}} +
b\ket{\bar{\varphi}_{2}\bar{\bar{\psi}}_{2}}\right) + \beta
\left(\ket{\phi_{2} \varphi_{2}\psi_{2}} +
\ket{\bar{\phi}_2\varphi_{2}\bar{\psi}_{2}}\right)$, with $a\cdot
b\neq 0$. Defining non-singular $F^{[k]}$ such that
$F^{[2]}[\ket{\phi_{2}}]=\ket{0}$,
$F^{[2]}[\ket{\bar{\phi}_{2}}]=\ket{1}$,
$F^{[3]}[\ket{\varphi_{2}}]=\ket{0}$,
$F^{[3]}[\frac{b}{a}\ket{\bar{\varphi}_{2}}]=\ket{1}$,
$F^{[4]}[\ket{\psi_{2}}]=\ket{0}$,
$F^{[4]}[\ket{\bar{\psi}_{2}}]=\ket{1}$ and $F^{[1]}$ as in
preceding cases, we obtain the canonical vector
$\ket{0\phi00}+\ket{0\phi1\psi}+\ket{1000}+\ket{1101}$, with the
restrictions that $\ket{\psi}\neq 0$ and $\ket{\phi}\neq \ket{0,1}$,
simultaneously (otherwise we would be reconsidering previous
classes). With the use of $\sigma_{x}$ upon qubit $4$, we are also
embracing the case in which
$\lin\{\ket{\Psi_{1}},\ket{\varphi_{2}\bar{\psi}_{2}}\} =
\lin\{00,00\}=\lin\{\ket{\varphi_{2} \bar{\psi}_{2}} ,
\ket{\bar{\varphi}_{2} \bar{\bar{\psi}}_{2}}\}$, where now the
double bar denotes linear independence with respect to
$\ket{\bar{\psi}_{2}}$.

\item[ii.] $\lin\{\ket{\Psi_{1}} , \ket{\varphi_{2} \psi_{2}}\} =
\lin\{00,\Psi\} = \lin\{\ket{\varphi_{2}\psi_{2}} ,
\ket{\varphi_{2}\bar{\bar{\psi}}_{2}} +
\ket{\bar{\varphi}_{2}\psi_{2}}\}$, with the same convention as
before for the double bar. Expressing $\ket{\Psi_{1}} =
a\ket{\varphi_{2}\psi_{2}} +
b\left(\ket{\varphi_{2}\bar{\bar{\psi}}_{2}} +
\ket{\bar{\varphi}_{2}\psi_{2}}\right)$ and with a similar argument
to the preceding case, we arrive at the canonical state
$\ket{0\phi0\psi}+\ket{0\phi10}+\ket{1000}+\ket{1101}$, with the
restrictions that $\ket{\psi}\neq 0$ and $\ket{\phi}\neq \ket{0,1}$,
simultaneously (otherwise we would be reconsidering previous
classes). By using $\sigma_{x}$ upon qubit $4$, we are also
considering the case
$\lin\{\ket{\Psi_{1}},\ket{\psi_{2}\bar{\phi}_{2}}\}=\lin\{00,\Psi\}$.
\end{enumerate}

\subsubsection{$\mathfrak{W}=\lin\{0_{1}\Psi_{23},0_{3}\Psi_{12}\}$}

Although this case can also be analyzed with similar arguments, we
will use permutation symmetry among qubits $2$ and $3$ in the case
$\lin\{0_{1}\Psi_{23},0_{2}\Psi_{13}\}$ to arrive at the canonical
states $\ket{0\phi00}+\ket{0\phi\varphi1}+\ket{1000}+\ket{1110}$ and
$\ket{0\phi\varphi0}+\ket{0\phi01}+\ket{1000}+\ket{1110}$.

\subsubsection{$\mathfrak{W}=\lin\{0_{2}\Psi_{13},0_{2}\Psi_{13}\}$}

Under permutation symmetry between qubits $2$ and $3$ in the case
$\lin\{0_{1}\Psi_{23},0{1}\Psi_{23}\}$, we rapidly find two
genuinely entangled canonical states, namely
$\ket{0000}+\ket{1010}+\lambda_{1}\ket{0101}+\lambda_{2}\ket{1111}$
and $\ket{0000} + \ket{1010} + \lambda_{1} \ket{0101} +
\lambda_{1}\ket{0001}+\lambda_{2}\ket{1011}+\lambda_{2}\ket{1110}$,
and one degenerate canonical state of type $\Psi_{13}\Psi_{24}$
(apart from the already considered).

\subsubsection{$\mathfrak{W}=\lin\{0_{2}\Psi_{13},0_{3}\Psi_{12}\}$}

Under permutation symmetry between qubits $2$ and $4$ in the case
$\lin\{0_{1}\Psi_{23},0_{3}\Psi_{12}\}$ we rapidly find two
genuinely entangled canonical states, namely
$\ket{000\psi}+\ket{0\phi1\psi}+\ket{1000}+\ket{1101}$ and
$\ket{0\phi0\psi}+\ket{001\psi}+\ket{1000}+\ket{1101}$.

\subsubsection{$\mathfrak{W}=\lin\{0_{3}\Psi_{12},0_{3}\Psi_{12}\}$}

Under permutation symmetry between qubits $2$ and $4$ in the case
$\lin\{0_{1}\Psi_{23},0_{1}\Psi_{23}\}$, we rapidly find two
genuinely entangled canonical states, namely
$\ket{0000}+\ket{1001}+\lambda_{1}\ket{0110}+\lambda_{2}\ket{1111}$
and
$\ket{0000}+\ket{1001}+\lambda_{1}\ket{0010}+\lambda_{1}\ket{0100} +
\lambda_{2}\ket{1011}+\lambda_{2}\ket{1101}$, and one degenerate
canonical state of type $\Psi_{14}\Psi_{23}$ (apart from the already
considered).

\subsubsection{$\mathfrak{W}=\lin\{0_{1}\Psi_{23},GHZ\}$}

The generic case will be of the form $\mathfrak{W} =
\lin\{\ket{\phi_{1}\Psi_{1}} , \ket{\phi_{2}\varphi_{2} \psi_{2}} +
\ket{\bar{\phi}_{2}\bar{\varphi}_{2}\bar{\psi}_{2}}\}$. Finding
non-singular transformations $F^{[i]}$ as before, we can write
$\mathfrak{W}=\lin\{\ket{\phi\Psi},\ket{000}+\ket{111}\}$. A generic
vector in $\mathfrak{W}$ will then be of the form
$\alpha\ket{\phi\Psi}+\beta\left(\ket{000}+\ket{111}\right)$. The
restrictions on $\ket{\phi}$ and $\ket{\Psi}$ (thus on
$\ket{\phi_{1}}$ and $\ket{\Psi_{1}}$) will be that no other
$0_{i_{k}}\Psi$ must belong to $\mathfrak{W}$. For instance, a
particular restriction will be that $\ket{\Psi}\notin\lin\{00,11\}$,
since otherwise $\ket{\Psi}=a\ket{00}+b\ket{11}$, $a\cdot b\neq0$
and then a generic state in $\mathfrak{W}$ will adopt the form
$\left(\alpha a\ket{\phi}+\beta\ket{0}\right)\ket{00}+\left(\alpha
b\ket{\phi}+\beta\ket{1}\right)\ket{11}$. It is always possible to
find non-null $\alpha,\beta$ such that the preceding state is a
$0_{1}\Psi_{23}$ vector different to $\ket{\phi\Psi}$, i.e.\ another
$0_{1}\Psi_{23}$ vector: we would be in the
$\lin\{0_{1}\Psi_{23},0_{1}\Psi_{23}\}$ case. The canonical state
will be given by $\ket{0\phi\Psi}+\ket{1000}+\ket{1111}$.

\subsubsection{$\mathfrak{W}=\lin\{0_{2}\Psi_{13},GHZ\}$}

By permutation symmetry between qubits $2$ and $4$, the canonical
state will be $\ket{0\Psi_{24}\phi_{3}}+\ket{1000}+\ket{1111}$,
where the first factor denotes an entangled state between qubits $2$
and $4$ and two factor states involving state $1$ and $3$.

\subsubsection{$\mathfrak{W}=\lin\{0_{3}\Psi_{12},GHZ\}$}

By permutation symmetry between qubits $2$ and $3$, the canonical
state will be $\ket{0\Psi_{23}\phi_{4}}+\ket{1000}+\ket{1111}$,
where the first factor denotes an entangled state between qubits $2$
and $3$ and two factor states involving state $1$ and $4$.

\subsubsection{$\mathfrak{W}=\lin\{0_{1}\Psi_{23},W\}$}

By $\mathfrak{W} = \lin\{0_{1}\Psi_{23},W\}$ it is meant that only
one degenerate tripartite vector of type $0_{1}\Psi_{23}$ belongs to
$\mathfrak{W}$, all the rest being of type $W$. It is immediate to
prove that this is impossible, i.e.\ that there is always possible
to find a $GHZ$ vector in $\mathfrak{W}$. The generic case will be
given by $\mathfrak{W}=\lin\{\ket{\phi_{1}
\varphi_{1}\psi_{1}},\ket{\phi_{2}\varphi_{2}\bar{\psi}_{2}} +
\ket{\phi_{2}\bar{\varphi}_{2}\psi_{2}}+\ket{\bar{\phi}_{2}
\varphi_{2} \psi_{2}}\}$. Defining $F^{[k]}$ so that $F^{[2]}
\ket{\psi_{2}} = \ket{0}$, $F^{[2]}\ket{\bar{\phi}_{2}}=\ket{1}$ and
similarly for $k=3,4$ and using the same $F^{[1]}$ as in preceding
cases, the coefficient matrix for a generic \footnote{Up to the
$0_{1\Psi}$ generator.} vector
$\ket{\phi\Psi}+\lambda\left(\ket{000}+\ket{111}\right)$ in
$\mathfrak{W}$ will be

\begin{equation}
\begin{pmatrix}
\phi_{0}\Psi_{00} & \phi_{0}\Psi_{01} + \lambda & \phi_{0} \Psi_{10}
+ \lambda & \phi_{0}\Psi_{11}\\
\phi_{1}\Psi_{00}+\lambda & \phi_{1}\Psi_{01} & \phi_{1}\Psi_{10} &
\phi_{1}\Psi_{11}
\end{pmatrix}
\end{equation}

Now there is a continuous range of values $\lambda\in\mathbb{C}$
such that $\rg(C^{(i)})=2$. Focus upon the set of which such that
$\textrm{det}W_{2}\neq 0$ (again a continuous range of them), then
it is clear that the spectrum of $W_{2}^{-1}W_{1}$ will be in
general non-degenerate, driving us to a $GHZ$ vector. Thus we are in
fact in the case $\lin\{0_{1}\Psi_{23},GHZ\}$.

\subsubsection{$\mathfrak{W}=\lin\{0_{2}\Psi_{13},W\}$}

By permutation symmetry the same analysis as in the preceding case
allows us to conclude that we are in the
$\lin\{0_{2}\Psi_{13},GHZ\}$ case.

\subsubsection{$\mathfrak{W}=\lin\{0_{3}\Psi_{12},W\}$}

By permutation symmetry the same analysis as in the preceding case
allows us to conclude that we are in the
$\lin\{0_{3}\Psi_{12},GHZ\}$ case.

\subsubsection{$\mathfrak{W}=\lin\{GHZ,W\}$}

In this class there exist no degenerate tripartite vectors, i.e.\
all of them are either of type $GHZ$ or of type $W$. With the same
arguments as before, the canonical state is found to be of the form
$\ket{1 \phi \varphi \psi} + \ket{1\bar{\phi}
\bar{\varphi}\bar{\psi}} + \ket{0001} + \ket{0010} + \ket{0100}$,
where the product vectors $\ket{\phi}$, $\ket{\bar{\phi}}$,
$\ket{\varphi}$, $\ket{\bar{\varphi}}$, $\ket{\psi}$ and
$\ket{\bar{\psi}}$ are restricted to produce no degenerate state in
$\mathfrak{W}$.

\subsubsection{$\mathfrak{W}=\lin\{GHZ,GHZ\}$}

The notation is meant to describe the situation in which only $GHZ$
vectors are contained in $\mathfrak{W}$. Apart for particular
instances in which a degenerate tripartite vector may appear as a
linear combination of both $GHZ$ generators (which we rule out,
since they have been already dealt with), we will show that there
will always exist at least one $W$ vector in $\mathfrak{W}$. The
generic case is given by
$\lin\{\ket{\phi_{1}\varphi_{1}\psi_{1}}+\ket{\bar{\phi}_{1}
\bar{\varphi}_{1}\bar{\psi}_{1}},\ket{\phi_{2}\varphi_{2}\psi_{2}} +
\ket{\bar{\phi}_{2}\bar{\varphi}_{2}\bar{\psi}_{2}}\}$. Then, after
defining non-singular transformations $F^{[k]}$  in a similar
fashion as in all preceding cases, the coefficient matrix for a
generic vector in $\mathfrak{W}$ will be given by

\begin{widetext}
\begin{equation}
\begin{pmatrix}
\phi_{0}\varphi_{0} \psi_{0} + \bar{\phi}_{0} \bar{\varphi}_{0}
\bar{\psi}_{0} + \lambda & \phi_{0}\varphi_{0}\psi_{1} +
\bar{\phi}_{0} \bar{\varphi}_{0} \bar{\psi}_{1} &
\phi_{0}\varphi_{1}\psi_{0} +
\bar{\phi}_{0}\bar{\varphi}_{1}\bar{\psi}_{0} & \phi_{0} \varphi_{1}
\psi_{1}+\bar{\phi}_{0}\bar{\varphi}_{1}\bar{\psi}_{1}\\
\phi_{1}\varphi_{0}\psi_{0} +
\bar{\phi}_{1}\bar{\varphi}_{0}\bar{\psi}_{0} &
\phi_{1}\varphi_{0}\psi_{1} +
\bar{\phi}_{1}\bar{\varphi}_{0}\bar{\psi}_{1} &
\phi_{1}\varphi_{1}\psi_{0}+\bar{\phi}_{1}\bar{\varphi}_{1}
\bar{\psi}_{0} &
\phi_{1}\varphi_{1}\psi_{1}+\bar{\phi}_{1}\bar{\varphi}_{1}
\bar{\psi}_{1}+\lambda
\end{pmatrix}
\end{equation}
\end{widetext}

There is a continuous range of vales $\lambda$ such that
$\rg(C^{(i)})=2$ for all $i=1,2,3$. Focus upon the set of which such
that $\rg(W_{1})=2$, then there will always exist a value
$\lambda^{*}$ such that $\sigma(W_{1}^{-1}W_{2})$ is degenerate
($\lambda^{*}$ is the solution to the equation
$\Delta(W_{1}^{-1}W_{2})=0$, where $\Delta(A)$ denotes the
discriminant of the characteristic polynomial of the matrix $A$). We
are then always in the case $\lin\{GHZ,W\}$.

\subsubsection{$\mathfrak{W}=\lin\{W,W\}$}

The notation is meant to describe the situation in which only $W$
vectors are contained in $\mathfrak{W}$. Apart for particular
instances in which a degenerate tripartite vector may appear as a
linear combination of both $W$ generators (which we rule out, since
they have been already dealt with), we will show that there will
always exist at least one $GHZ$ vector in $\mathfrak{W}$. The
generic case is given by $\lin\{\ket{\phi_{1}
\varphi_{1}\bar{\psi}_{1}} + \ket{\phi_{1}\bar{\varphi}_{1}\psi_{1}}
, \ket{\bar{\phi}_{1} \varphi_{1}\psi_{1}} ,
\ket{\phi_{2}\varphi_{2}\bar{\psi}_{2}} +
\ket{\phi_{2}\bar{\varphi}_{2}\psi_{2}},\ket{\bar{\phi}_{2}
\varphi_{2}\psi_{2}}\}$. Then, after defining non-singular
transformations $F^{[k]}$  in a similar fashion as in all preceding
cases, the coefficient matrix for a generic vector in $\mathfrak{W}$
will be given by

\begin{widetext}
\begin{equation}
\left(\begin{smallmatrix} \phi_{0}\varphi_{0}\bar{\psi}_{0} +
\phi_{0} \bar{\varphi}_{0} \psi_{0} +
\bar{\phi}_{0}\varphi_{0}\psi_{0} &
\phi_{0}\varphi_{0}\bar{\psi}_{1} +
\phi_{0}\bar{\varphi}_{0}\psi_{1} + \bar{\phi}_{0}
\varphi_{0}\psi_{1} + \lambda & \phi_{0}\varphi_{1}\bar{\psi}_{0} +
\phi_{0}\bar{\varphi}_{1} \psi_{0} +
\bar{\phi}_{0}\varphi_{1}\psi_{0} + \lambda & \phi_{0}
\varphi_{1}\bar{\psi}_{1}+\phi_{0}\bar{\varphi}_{1}\psi_{1} +
\bar{\phi}_{0}
\varphi_{1}\psi_{1}\\
\phi_{1}\varphi_{0}\bar{\psi}_{0} + \phi_{1}\bar{\varphi}_{0}
\psi_{0}+\bar{\phi}_{1}\varphi_{0}\psi_{0}+\lambda &
\phi_{1}\varphi_{0} \bar{\psi}_{1}+\phi_{1}\bar{\varphi}_{0}
\psi_{1}+\bar{\phi}_{1}\varphi_{0}\psi_{1} &
\phi_{1}\varphi_{1}\bar{\psi}_{0} +
\phi_{1}\bar{\varphi}_{1}\psi_{0}+\bar{\phi}_{1}\varphi_{1}\psi_{0}
& \phi_{1}\varphi_{1}\bar{\psi}_{1} +
\phi_{1}\bar{\varphi}_{1}\psi_{1}+\bar{\phi}_{1}\varphi_{1}\psi_{1}
\end{smallmatrix}\right)
\end{equation}
\end{widetext}

There is a continuous range of vales $\lambda$ such that
$\rg(C^{(i)})=2$ for all $i=1,2,3$. Focus upon the set of which such
that $\rg(W_{2})=2$, then there will always exist values
$\lambda^{*}$ such that $\sigma(W_{1}^{-1}W_{2})$ is non-degenerate
($\lambda^{*}$ is not the solution to the equation
$\Delta(W_{2}^{-1}W_{1})=0$, where $\Delta(A)$ denotes the
discriminant of the characteristic polynomial of the matrix $A$). We
are then always in the case $\lin\{GHZ,W\}$, too.

\subsection{Counting entanglement classes}

The preceding classification of states under the SLOCC criterion
reveals a block structure for the entire state space
$\left(\mathbb{C}^{2}\right)^{\otimes4}$, with a continuous range of
non-equivalent vectors inside each block (except for some of them).
Considering each block as an entanglement class itself, then we have
found $18$ degenerate and $16$ genuine classes, i.e.\ a total of
$34$ entanglement classes.

Furthemore, taking into account permutation symmetry among the four
qubits, indeed we have proven that \emph{four qubits can be
entangled in $8$ inequivalent ways}. This is in apparent
contradiction to the result found in \cite{VerDehMooVer02a}, where
four qubits are claimed to be entangled in $9$ possible ways. A
closer look at this work reveals that one of the blocks in that
classification contains no genuinely entangled vector, namely the
class $L_{0_{3\oplus\bar{1}}0_{3\oplus 1}}$ with canonical state
$\ket{0000}+\ket{0111}$. We would like to express our conviction
that this difference stems from a distinct way of counting or
considering entanglement classes. We believe ours shows a greater
agreement with the philosophy of the seminal paper by D\"{u}r
\emph{et al}~\cite{DurVidCir00a}. Thus we have proven the following

\begin{thm}\label{MaiThe}
Let $\ket{\Phi}$ be a genuinely entangled $4$-partite pure state.
Then $\ket{\Phi}$ belongs to one of the following $8$ genuine
entanglement classes and is equivalent to some of its canonical
vectors:

\begin{widetext}
\begin{center}
\begin{table}[h!]
\begin{center}
\begin{tabular}{|c|l|c|c|}\hline
\textbf{Class ($\mathfrak{W}$)}
&\;\;\;\;\;\;\;\;\;\;\;\;\;\;\;\;\;\;\;\;\;\;\;\;\;\;\;\;\;\;\textbf{Canonical
States} & \textbf{Name} & \textbf{Proposed Notation}\\\hline
&&&\\
$\lin\{000,000\}$ & $\ket{0000}+\ket{1111}$ & $GHZ$ &
$\mathfrak{W}_{000,000}$\\
&&&\\\hline
&&&\\
$\lin\{000,0_{k}\Psi\}$ & $\ket{0000}+\ket{1100}+\ket{1111}$
& - & $\mathfrak{W}_{000,0_{k}\Psi}$\\
                        & $\ket{0000} + \ket{1101} +
                        \ket{1110}$ &  & \\
&&&\\\hline
&&&\\
$\lin\{000,GHZ\}$ & $\ket{0\phi\varphi\psi} + \ket{1000} +
\ket{1111}$
& - &  $\mathfrak{W}_{000,GHZ}$\\
&&&\\\hline
&&&\\
$\lin\{000,W\}$   & $\ket{1000} + \ket{0100} + \ket{0010} +
\ket{0001}$
& $W$  & $\mathfrak{W}_{000,W}$\\
&&&\\\hline
&&&\\
$\lin\{0_{k}\Psi,0_{k}\Psi\}$ & $\ket{0000} + \ket{1100} +
\lambda_{1} \ket{0011}+\lambda_{2}\ket{1111}$ & $\Phi_{4}$  &
$\mathfrak{W}_{0_{k} \Psi,0_{k}\Psi}$ \\ & $\ket{0000} + \ket{1100}
+ \lambda_{1} \ket{0001} + \lambda_{1}\ket{0010}+\lambda_{2}
\ket{1101}+\lambda_{2}\ket{1110}$ & &\\
&&&\\\hline
&&&\\
$\lin\{0_{i}\Psi,0_{j}\Psi\}$ & $\ket{0\phi00} + \ket{0 \phi1 \psi}
+ \ket{1000}+\ket{1101}$ & - &  $\mathfrak{W}_{0_{i}\Psi,0_{j}
\Psi}$
\\ & $\ket{0\phi0\psi} + \ket{0\phi10} + \ket{1000}
+ \ket{1101}$ & &\\
&&&\\\hline
&&&\\
$\lin\{0_{k}\Psi,GHZ\}$ & $\ket{0\phi\Psi} + \ket{1000} +
\ket{1111}$ & - & $\mathfrak{W}_{0_{k}\Psi,GHZ}$ \\
&&&\\\hline
&&&\\
$\lin\{GHZ,W\}$ & $\ket{0001} + \ket{0010}+\ket{0100} +
\ket{1\phi\varphi \psi} + \ket{1\bar{\phi}\bar{\varphi}
\bar{\psi}}$& - & $\mathfrak{W}_{GHZ,W}$ \\
&&&\\\hline
\end{tabular}
\end{center}
\caption{Genuine entanglement classes for four qubits\label{Table}}
\end{table}
\end{center}
\end{widetext}
The canonical vectors obtained by permutation among qubits $2,3,4$
have been omitted in each entry. Restrictions upon the range of the
free parameters in each class amounts to not producing a precedent
structure for the right singular subspace $\mathfrak{W}$.

\end{thm}

Names should be given to all classes. A systematic notation for
these classes could possibly be $\mathfrak{W}_{\Psi_{i},\Psi_{j}}$,
where the subindices indicate the generators of $\mathfrak{W}$. In
this convention $GHZ=\mathfrak{W}_{000,000}$,
$W=\mathfrak{W}_{000,W}$ and
$\Phi_{4}=\mathfrak{W}_{0_{1}\Psi,0_{1}\Psi}$. However a further
discrimination referring to the particular canonical state must be
explicitly stated.

\section{Concluding remarks}
\label{ConRem}

We have given the classification of entangled states of four qubits
under stochastic local operations and classical communication using
the inductive approach introduced in \cite{LamLeoSalSol06a}. This
approach is based on the analysis of the structure of the right
singular subspace $\mathfrak{W}$ of the coefficient matrix of the
state in a product basis according to the entanglement classes for
three qubits.

Agreeing to consider each structure of $\mathfrak{W}$ as a single
entanglement class, we have found $18$ degenerate and $16$ genuine
classes (totally $34$ classes), where permutation is explicitly
included in the counting. Taking into account the permutation among
the qubits, $8$ genuine classes can be considered, compiled in table
\ref{Table}. As expected, in most of the classes a continuous range
of strictly non-equivalent states is contained, although with
similar structure.

This result allows us to predict that there will be at most $765$
entanglement classes (permutation included) for $5$-partite systems,
$595$ at most genuine and $170$ at most degenerate (cf.\
\cite{LamLeoSalSol06a}). No doubt, this classification stands up as
a formidable task that our inductive method may help to elucidate.

\section*{Acknowledgments}

LL and JL acknowledge financial support from the Spanish MEC
projects No. FIS2005-05304, DS from No. FIS2004-01576, and ES from
EU RESQ, EuroSQIP, and DFG SFB631 projects. LL also acknowledges
support from the FPU grant No. AP2003-0014.

\end{document}